\documentclass[prodmode,acmtoms]{acmsmall} % Aptara syntax

\usepackage[T1]{fontenc}

\usepackage{amsmath}	% Grundlegende Mathe Bibliothek
\usepackage{listings}	% source code darstellen
\usepackage{booktabs}
\lstdefinestyle{normal}{ %
language=C++,                % the language of the code
numbers=none,
basicstyle=\footnotesize,       % the size of the fonts that are used for the code
frame=single,                   % adds a frame around the code
tabsize=2,                      % sets default tabsize to 2 spaces
captionpos=b,                   % sets the caption-position to bottom
breaklines=true,                % sets automatic line breaking
breakatwhitespace=true,        % sets if automatic breaks should only happen at whitespace
}

\lstdefinestyle{numbers}{
language=C++,                % the language of the code
numbers=left,
basicstyle=\footnotesize,       % the size of the fonts that are used for the code
frame=single,                   % adds a frame around the code
tabsize=2,                      % sets default tabsize to 2 spaces
captionpos=b,                   % sets the caption-position to bottom
breaklines=true,                % sets automatic line breaking
breakatwhitespace=true        % sets if automatic breaks should only happen at whitespace
}
\usepackage{subfig}

\usepackage{tikz}
\usetikzlibrary{arrows, decorations, matrix}
\usetikzlibrary{shapes.multipart}
\usetikzlibrary{positioning}
\usetikzlibrary{fit}
\usetikzlibrary{shapes.geometric,backgrounds,calc}

\newcommand{\R}{\mathbb{R}}
\newcommand{\aeq}{\kern.35em\text{\small+}\kern-.35em=}
\newcommand{\eqdot}{\ .}

\usepackage[ruled]{algorithm2e}

\SetAlFnt{\small}
\SetAlCapFnt{\small}
\SetAlCapNameFnt{\small}
\SetAlCapHSkip{0pt}
\IncMargin{-\parindent}

\acmVolume{V}
\acmNumber{N}
\acmArticle{A}
\acmYear{YYYY}
\acmMonth{0}

\doi{0000001.0000001}

\issn{1234-56789}

\begin{document}

\markboth{M. Sagebaum et al.}{High-Performance Derivative Computations using CoDiPack}

\title{High-Performance Derivative Computations using CoDiPack}
\author{
MAX SAGEBAUM
\affil{Chair for Scientific Computing, TU Kaiserslautern}
TIM ALBRING
\affil{Chair for Scientific Computing, TU Kaiserslautern}
NICOLAS R. GAUGER
\affil{Chair for Scientific Computing, TU Kaiserslautern}}

\begin{abstract}
There are several AD tools available, which all implement different strategies for the reverse mode of AD.
The major strategies are primal value taping (implemented e.g. by ADOL-c) and Jacobi taping (implemented e.g. by adept and dco/c++).
Especially for Jacobi taping, recent advances by using expression templates make this approach very attractive for large scale software.
The current implementations are either closed source or miss essential features and flexibility.
Therefore, we present the new AD tool CoDiPack (Code Differentiation Package) in this paper.
It is specifically designed for a minimal memory consumption and optimal runtime, such that it can be used for the differentiation of large scale software.
An essential part of the design of CoDiPack is the modular layout and the recursive data structures, which do not only allow the efficient implementation of the Jacobi taping approach, but will also enable other approaches like the primal value taping or new research ideas.
We will also present the performance value of CoDiPack on a generic PDE example and on the SU2 code.
\end{abstract}

%TODO



%TODO

\acmformat{Max Sagebaum, Tim Albring, Nicolas R. Gauger, 2016. CoDiPack Paper.}

\begin{bottomstuff}
Author's addresses: M. Sagebaum {and} T. Albring {and} N. R. Gauger, Chair for Scientific Computing, TU Kaiserslautern
\end{bottomstuff}

\maketitle

\section{Introduction}

Algorithmic Differentiation (AD) describes the mathematical theory how a computer program can be differentiated.
A basic introduction to AD will be given in the second section of this paper.
However, we already state here the fundamental result of the reverse mode of AD, namely, that for each statement $y = w(x)$ with $w : \R^n \rightarrow \R^m$ in the software the corresponding adjoint statement
\begin{equation}
	\label{eq.revADInt}
	\bar x \aeq \frac{d w}{d x}^T \bar y; \quad \bar y = 0
\end{equation}
must be evaluated. When the software is written in Fortan, source-code transformation \cite{andreas_book} is the method of choice to automate the generation of code to compute \eqref{eq.revADInt}. The application of source-code transformation to C++ code is limited due to the complex language structure. Instead, Operator Overloading (OO) has been proven to be more appropriate. All of the available tools that implement AD based on the OO approach, have the problem of storing the data and evaluating equation \eqref{eq.revADInt} in a number of different ways depending on the programming language, field of application or just the personal preference. However, they all have in common, that since the flow of data is reversed, it is required to either store the values of $x$ or to directly store the Jacobian $\frac{d w}{d x}$ during the evaluation. This corresponds to what we refer to as the \emph{Primal taping} or the \emph{Jacobian taping} method, respectively.

The most well known representative that implements the former method is \emph{ADOL-C} \cite{walther2012getting}. It is released under the EPL and GPL2 license and is widely used in science for small and medium sized applications. Its shortcomings are the runtime and memory overhead when applied to large-scale problems. Still, its level of maturity and robustness makes it an excellent aid for validation and verification, when developing new operator overloading AD tools.

The \emph{Jacobian taping} method is quite popular in combination with expression templates (ET) \cite{Veldhuizen1995,Aubert2001} to increase the runtime efficiency. Two examples of tools that make use of this approach are for example \emph{adept} \cite{hogan2014fast} and \emph{dco} \cite{dcoPaper}. \emph{adept} is released as open-source and offers high performance and a relatively low memory footprint, but it lacks important features like higher-order differentiation, a vector-mode or a (tapeless) forward mode implementation. Although \emph{dco} offers all these features, we think that its proprietary license makes it unattractive for individuals and organizations in science and industry to use it in their in-house or open-source software. Furthermore, in our opinion, both tools have in common that their structure is quite inflexible so that they are not easily extensible. These were the initial reasons for starting the development of the \emph{Code Differentiation Package} (CoDiPack). Table \ref{tab.overview} shows a comparison in terms of taping methods, implementation and license of the tools mentioned above.

\begin{table}[h]
	\centering
	\tbl{Overview of AD Tools for C++ (non-exhaustive list)\label{tab.overview}}{ 
		\begin{tabular}{c|c|c|c|c}
		Tool & Taping Method(s) & Implementation & License\\ \hline 
		\textbf{CoDiPack} & Jacobian, Primal$^{(*)}$ & OO/ET & Open-Source (GPL3) \\
		ADOL-C & Primal & OO & Open-Source (EPL/GPL2)\\
		adept & Jacobian & OO/ET & Open-Source (Apache2.0) \\
		dco & Jacobian & OO/ET & proprietary  \\
		
		\end{tabular}
	} 
	\footnotesize{$^{(*)}$as of version 1.3, not covered in this work}
\end{table}
Today, the main focus of \emph{CoDiPack} is the application of AD to high-performance computing (HPC) and industrial-grade software, while still maintaining its open-source philosophy. Hence, the following properties are always considered during the development:
\subsubsection*{Efficiency}
One of the top priorities is to write an efficient tool for the application of AD and to reduce the runtime overhead of the differentiated software.
Similarly, the memory should be used as efficient as possible and the layout of the memory access should conform to modern HPC standards.
\emph{CoDiPack} offers a number of different ways such that the characteristics of the code base on which it is applied can be maintained.
This will improve the performance of the differentiated software.
Examples are the support for c-like memory operations (e.g. memcpy) or a data layout that is optimized for running several reverse sweeps in a row (see section 2).
\subsubsection*{Usability}
For the tool to be accepted by the users it is important that the time to get familiar with the API is quite short.
Thus first results should be obtained in a minimal amount of time.
This is accomplished by having distinct and well documented user interfaces as well as simple tutorials.
Furthermore, to apply \emph{CoDiPack} usually no major changes in the software are required.

\subsubsection*{Extensibility}
To increase the pace in AD related research, \emph{CoDiPack} is designed in a modular way such that new features (e.g. new taping strategies), high-level operations or the memory layout can be easily modified or added. \\

This paper serves the purpose to highlight in more detail the structure of the implementation and show how it affects the resulting performance.
First an introduction to AD is given in section \ref{sec.ad}, which is followed by an explanation of expression templates in section \ref{sec.expr}.
The modular approach of the CoDiPack implementation is then described in section \ref{sec.mod}.
Here, the basic idea for the efficient application of AD through operator overloading is described and the design choices for the efficient and extensible memory structures are explained.
The performance is investigated by applying \emph{CoDiPack} to a generic PDE example and to the open-source computational fluid dynamics suite SU2 \cite{economon2015su2} in section \ref{sec.test}.

\section{Algorithmic Differentiation}\label{sec.ad}

The part in each software for which the derivatives are needed can be described in all cases as a function with the header \emph{void func(const double x[], double y[])}.
In most cases the input variables $x$ and the output variables $y$ will be not straight arrays but collections of several variables.
These variables can also be members of structures that are in a local or global scope.
In an extreme case the header for \emph{func} will be just \emph{void func(void)}, but internally \emph{func} accesses millions of input and output variables from the global scope.
Nevertheless, \emph{func} can always be described as a the mathematical function $f: \R^n \rightarrow \R^m$ with 
\begin{equation}
y = f(x).
\end{equation} 
We assume that the numerical evaluation of $f$ can be represented as a sequence of $l$ statements $\varphi_i: \mathbb{R}^{n_i} \mapsto \mathbb{R}$.
Each of this statements represents a local evaluation procedure with an arbitrary complex right-hand side, for example
\begin{eqnarray}\label{eq.ex_statement}
v_4 = \varphi_4(v_1,v_2,v_3) = -10*v_2*\exp(v_3) + \ln(v_1) - 3 *10^{7} * v_3 * (v_2-1) * \sqrt{v_1}.
\end{eqnarray}
It is then possible to write any numerical evaluation in an arbitrary computer program using the general procedure shown in Table \ref{tab.gen_eval_prod}.
\renewcommand{\arraystretch}{1.3} 
\begin{table}[h]
	\centering
		\caption{Evaluation procedure for a function $f$.}
	\begin{tabular}{llll}
		\toprule$v_{i}$ &$= x_i,$ &$i =$ & $1 \hdots n $ \\
		$v_{i + n}$ &$= \varphi_i(u_i),  $&$i =$ & $1  \hdots l$ \\
		$y_{i}$ &$= v_{n+l-i+1} ,$ &$i = $& $1 \hdots m$\\
		\hline
	\end{tabular}
	\label{tab.gen_eval_prod}
\end{table}
We have $u_i := \varphi(v_j)_{j \prec i} \in \mathbb{R}^{n_i}$, where the precedence relation $j \prec i$ means that variable $v_i$ depends directly on variable $v_j$.
That is, the vector $u_i$ is the concatenation of the $v_j$ on which $\varphi_i$ depends.
It is important to note, that the mathematical representation of $f$ and the implementation \emph{func} is not a one to one relation.
$f$ represents all the statements that are evaluated during one call of \emph{func} which implies, that e.g. loops are unrolled in $f$.
For large applications the number of statements $l$ can become quite large and can vary for different configurations.

The representation of $f$ can also be written it as the composition
\begin{equation} \label{eq:compf}
f(x) = Q_n \circ \Phi_l \circ \Phi_{l-1} \circ ... \circ \Phi_2 \circ \Phi_1 \circ P_n^T(x) \eqdot
\end{equation}
where the state transformation $\Phi_i: V \mapsto V$ sets $v_i$ to $\varphi_i(v_j)_{j \prec i}$ and keeps all other components $v_j$ for $i \not= j$ unchanged.
$V := \R^n \times \R^l$ is the state space of the evaluation of \emph{func}.
It contains the input values $x$, the intermediate values and the output values $y$ as the last $m$ values of the intermediate values.
$P_y \in \mathbb{R}^{n \times (n +l) } $ and $I_x \in \mathbb{R}^{m \times (n +l)}$ are the matrices that project an $(n+l)$-vector onto its first $n$ and last $m$ components, respectively.
By applying the chain rule for differentiation we get
\begin{equation} \label{eq.forward_relation}
\dot{y} = Q_m A_l A_{l-1} \hdots A_2 A_1 P_n^T \dot{x},
\end{equation}
where $A_i := \nabla \Phi_i$.
Using the analytic representation of $A_i$ \cite{andreas_book}, it is possible to write the tangent relation in equation \eqref{eq.forward_relation} as the evaluation procedure shown in table \ref{tab.ad_fwd}.
The matrix vector products are calculated in the same order as for the evaluation procedure in table \ref{tab.gen_eval_prod} and can be computed alongside the primal evaluation.
It is then possible to compute the matrix-vector product of the Jacobian and an arbitrary direction $\dot x$, i.e. equation \eqref{eq.forward_relation}, by evaluating the tangent interpretation.
\begin{table}[h]
	\centering
	\caption{Tangent Interpretation (Forward mode of AD).}
	\begin{tabular}{llll}
			\toprule$\dot v_{i}$ &$= \dot x_i,$ &$i =$ & $1 \hdots n $ \\
			$\dot v_{i + n}$ &$= \sum\limits_{j \prec i} \frac{\partial }{\partial v_j} \varphi_i(u_i) \bar v_{j},  $&$i =$ & $1  \hdots l$ \\
			$\dot y_{i}$ &$= \dot v_{n+l-i+1} ,$ &$i = $& $1 \hdots m$\\
			\hline
		\end{tabular}
	\label{tab.ad_fwd}
\end{table}

The tangent relation in equation \eqref{eq.forward_relation} gives also the alternative representation of the Jacobian of $f$ which can be written as
\begin{eqnarray}
\frac{\mathrm{d} f(x)}{\mathrm{d}x} =Q_m A_l A_{l-1} \hdots A_2 A_1 P_n^T.
\end{eqnarray}
By transposing the product we obtain the adjoint relation
\begin{equation} \label{eq.adjoint_relation}
\bar{x} = P_n A_1^T A_2^T \hdots A_{l-1}^T A_l^T Q_m^T \bar{y} =\left( \frac{\mathrm{d}f}{\mathrm{d}x} \right)^T \bar{y}.
\end{equation}
We can again use the analytic representation of $A_i$ \cite{andreas_book} to write the adjoint relation as the evaluation procedure shown in table \ref{tab.incr_ad_rec}.
All matrix-vector products are calculated for $i = l,l-1,\hdots, 1$ thus we have to go in reverse through the sequence of statements in Table \ref{tab.gen_eval_prod}.
We can then get the matrix-vector product involving the transposed of the Jacobian and an arbitrary seed vector $\bar{y}$, i.e. equation \eqref{eq.adjoint_relation}, by evaluating this adjoint interpretation.
\begin{table}[h]
	\centering
	\caption{Adjoint Interpretation (Reverse mode of AD).}
	\begin{tabular}{lll}
		%		\hline   $v_{i-n}$ & $= x_i,$ & $i=$  & $1 \hdots n$ \\
		%		$v_{i} $&   $= \varphi_i(v_j)_{j \prec i},$  &$i =$ & $1  \hdots l$ \\
		%		$y_{m-i}$ &$= v_{l-i} , $&$i =$ & $m-1 \hdots 0$ \\
		\toprule $\bar{v}_{n+l-i+1}$ &$= \bar{y}_{i}$ &$i = m \hdots 1$ \\
		$\bar{v}_{j}$ &$= \bar{v}_j + \bar{v}_{i+n} \cdot \frac{ \partial }{\partial v_j} \varphi_i(u_i), \  \bar{v}_i = 0 $  &$ j \prec i, i = l \hdots  1$ \\
		$\bar{x}_i $&$= \bar{v}_{i}$ &$i = n \hdots 1$ \\
		\bottomrule
	\end{tabular}
	\label{tab.incr_ad_rec}
\end{table}

The Jacobi taping approach, that we want to implement in this paper, will store the gradient $\frac{ \partial }{\partial v_j}\varphi_i(u_i), j \prec i$   of each statement.
An efficient handling of the statements and how the gradient can be computed will be discussed in the next section.
How the data is stored, is then explained in the CoDiPack implementation.

\section{Expression templates} \label{sec.expr}
One big issue for an efficient operator overloading AD tool implementation is that C++ only supports the overloading of unary and binary operators.
A regular implementation that follows the layout $Real \circ Real \rightarrow Real$ where $\text{Real}$ is the overloaded type would split the statement 
\begin{equation}
	w = ((a + b) * (c - d))^2
	\label{eq.stmtExample}
\end{equation}
 into
\begin{align*}
  t_1 & = a + b\\
  t_2 & = c - d\\
  t_3 & = t_1 * t_2\\
  w & = t_3^2
\end{align*}
which needs to store 7 Jacobies and 4 statements in total.
But the original statement would only need 4 Jacobies with 1 statement.
It is therefore much more efficient to treat the whole statement at once.

In order to achieve this, the expression template technique is used.
We change the layout of the operator to $Expr_A \circ Expr_B \rightarrow Expr_{A \circ B}$ where $Expr$ is some class that stores information about the operation that is performed.
It is very important how this information is stored.
If a regular inheritance scheme is used, then virtual functions calls would be required in every statement of the program.
When virtual function calls are used in high level data structures, their performance impact can be neglected, because they will occur not very often.
The situation changes, when they are used in every statement, then they are called millions of times in a second and the cpu is just loading the addresses for the virtual function.
Hence, we use expression templates to avoid the need of virtual functions.

The basic idea of expression templates is to define a template class that gets the extending class as a template argument.
Figure \ref{lst.exprImplementation} shows this in an example.
\begin{figure}
\begin{lstlisting}[style=numbers]
template<typename A>
class Expression {
  A& cast() { return static_cast<A&>(*this);}
  double value() { return this->cast().value();}
};

template<typename A, typename B>
class MULT : public Expression<MULT<A, B> > {
  const A& a;
  const B& b;
  
  MULT(const A& a, const B& b) : a(a), b(b) {}
  double value() { return a.value() * b.value();}
}
\end{lstlisting}
\caption{Expression template example.}
\label{lst.exprImplementation}
\end{figure}
The base class provides a cast method in line 3 to return a reference to the extending class.
Every method in the base class uses this method to call the implementation of the extending class.
In line 8 the structure \emph{MULT} is implemented and the \emph{Expression} interface is used as a base class.
If the type for the variable is called \emph{Real} then the statement from \eqref{eq.stmtExample} is represented by the structure
\begin{equation*}
  POW< MULT< ADD< Real, Real>, SUB< Real, Real> > > \eqdot
\end{equation*}
It represents the full statement and AD can use it to perform arbitrary computation on the whole statement.

For a Jacobi taping approach it is now necessary to compute the derivatives of the statement and the \emph{Expression} interface from figure \ref{lst.exprImplementation} needs to be extended in order to provide this information.
The implementation will be the AD reverse mode of the \emph{value} method in the \emph{MULT} expression example.
The \emph{getValue} method can be viewed as the algorithm
\begin{equation}
	\begin{aligned}
		a = & A(p) \\
		b = & B(q) \\
		w = & a * b ,
	\end{aligned}
\end{equation}
where $A$ and $B$ represent the expressions of the two arguments from the multiplication.
$p \in \R^s$ and $q \in \R^t$ represent the variables that are used in the expressions.
If reverse AD is applied to this algorithm, then the resulting algorithm is
\begin{equation}
	\begin{aligned}
		\bar a = & b * \bar w \\
		\bar b = & a * \bar w \\
		\bar q = & \frac{d B}{d q} * \bar b \\
		\bar p = & \frac{d A}{d p} * \bar a.
	\end{aligned}
	\label{eq.revBinary}
\end{equation}
The reverse algorithm states, that the Jacobies with respect to the arguments need to be computed and then the adjoint values are given recursively to the arguments.
Therefore, we can extend the expression interface with a new method \emph{calcGradient} that implements this algorithm.
Figure \ref{lst.calcJacImplemenation} shows a general implementation for an arbitrary binary operator.
The two methods, \emph{derivativeA} and \emph{derivativeB}, calculate the derivatives with respect to the first and second argument, they evaluate the first two lines in algorithm \eqref{eq.revBinary}.
For the last two lines, it is now necessary to do recursive calls of \emph{calcGradient} on the arguments of the expression.
The recursion is terminated in the variables of the expression, there the \emph{multiplier} argument of the method will contain the derivative of the whole expression with respect to this argument.
\begin{figure}
\begin{lstlisting}[style=numbers]
void calcGradient(const double& multiplier) {
  double dw_da = derivativeA(a.value(), b.value(), this->value()) * multiplier;
  double dw_da = derivativeB(a.value(), b.value(), this->value()) * multiplier;
  
  a.calcGradient(dw_da);
  b.calcGradient(dw_db);
}
\end{lstlisting}
\caption{Implementation for the Jacobi computation in an expression template.}
\label{lst.calcJacImplemenation}
\end{figure}
For the example statement \eqref{eq.stmtExample} a call of \emph{calcGradient(1.0)} on the expression template will evaluate the code in figure \ref{lst.exampleStmts}.
This should also be the code, that is generated by the compiler, after everything is inlined.

 ``generate'' the code in figure \ref{lst.exampleStmts}, because the compiler can see and inline all the functions of the expressions.
\begin{figure}
\begin{lstlisting}[style=numbers]
  double add = a.value() + b.value();
  double sub = c.value() - d.value();
  double mul = add * sub;
  double w = pow(mul, 2.0);
  
  double w_b = multiplier; // = 1.0
  double mul_b = 2 * mul * w_b;
  double sub_b = add * mul_b;
  double add_b = sub * mul_b;
  
  c.calcGradient(sub_b);
  d.calcGradient(-sub_b);
  a.calcGradient(add_b);
  b.calcGradient(add_b);
\end{lstlisting}
\caption{The code that a compiler should generate when \emph{calcGradient} is called on statement \eqref{eq.stmtExample}.}
\label{lst.exampleStmts}
\end{figure}
The remaining calls to \emph{calcGradient} are then implemented such that they access the tape and store the data, which will be covered in the next section.

\section{Design and layout of CoDiPack}\label{sec.mod}

As it is stated in the introduction, one of the main goals of CoDiPack is to be as fast and memory efficient as possible in order to be able to use CoDiPack in HPC environments.
Therefore, it needs to be carefully considered which data is stored and how this is done.

Which data is required, can be analyzed with the help of the reverse AD equation for a statement, as it is shown in table \ref{tab.incr_ad_rec}:
\begin{equation}
\bar{v}_{j} = \bar{v}_j + \bar{v}_{i+n} \cdot \frac{ \partial }{\partial v_j} \varphi_i(u_i), \  \bar{v}_i = 0 \quad j \prec i, i = l \hdots  1 \quad.
\label{eq.revOperator}
\end{equation}
In order to evaluate the Jacobi taping approach for this equation, the required data are the Jacobi $\frac{d \varphi}{d v}$ and the means to access the adjoint variables for $u$ and $v_j$.

The adjoint variables could be directly stored in the overloaded type, however, this is not a feasible solution since this information would be deleted when they run out of scope.
It is a common practice to use the identification, that is provided by the AD theory.
When the function $f$ is separated into the elemental operations $\phi_i$, every operation gets an index $i$ that runs from one to the maximum number of operations $l$.
Each variable can therefore be identified with the index of the elemental operation.
This index is implemented as a global counter, which is incremented each time a statement is stored in the AD tool and the current value of the counter is used as the index for the value on the left hand side of the assignment.

The memory for the data of one elemental operation with $k$ input values is therefore $k$ double values for the Jacobi, $k$ indices for the arguments and one index for the output value.
In addition one byte is needed to store the number of arguments.
This assumes that a statement has no more than 255 arguments, which is a reasonable assumption for normal code.
The total memory requirement for each statement is then $k * 8 + (k + 1)* 4 + 1 = 12 * k + 5$ bytes, but this not yet optimal.

The indexing scheme that we use increases the index of the left hand side by one for each statement and the index is stored directly for the reverse evaluation.
During the reverse evaluation all statements are evaluated in the exact same order but just reversed.
The index of the left hand can therefore be computed by decrementing it one by one and it is no longer necessary to store the index of the left hand side.
This reduces the memory requirement by $4$ bytes, which is then  $k * 8 + k* 4 + 1 = 12 * k + 1$ for each statement.
We call this indexing scheme ``linear indexing'', which is also used by dco \cite{dcoPaper}.

The next subsections will now introduce the efficient computation and storing of the required data for the Jacobi taping.

\subsection{Design and implementation of the expression templates}
The computation of the Jacobi for the elemental operation is described in the expression template implementation.
The CoDiPack interface for them is shown in figure \ref{lst.expressionInterface}.
\begin{figure}
\begin{lstlisting}[style=numbers]
  template<typename R, class A>
  struct Expression {

    static const bool storeAsReference;
    typedef typename TypeTraits<R>::PassiveReal PassiveReal;

    inline const A& cast() const {
      return static_cast<const A&>(*this);
    }

    template<typename Data>
    inline void calcGradient(Data& data) const {
      cast().calcGradient(data);
    }

    template<typename Data>
    inline void calcGradient(Data& data, const R& multiplier) const {
      cast().calcGradient(data, multiplier);
    }

    inline const R getValue() const {
      return cast().getValue();
    }

  private:
    Expression& operator=(const Expression&) = delete;
  };
\end{lstlisting}
\caption{The interface definition for the expression templates in CoDiPack.}
\label{lst.expressionInterface}
\end{figure}
The \emph{cast} method and the \emph{getValue} method are the same as in figure \ref{lst.exprImplementation}, but an additional template parameter for the class is introduced.
It removes the constraint that only double types can be used in the computation, such that for example higher order derivatives can be computed.
The \emph{calcGradient} method is also extended by a template parameter, which defines some user data that can be used in the computations.
Furthermore, two versions of \emph{calcGradient} are defined, a two argument version and a version with one argument which assumes that the multiplier is equal to $1.0$.
This addition to the interface is made in order to give the compiler as much information as possible and to prevent unnecessary multiplications with 1.0.

The implementation of the interface for the unary operations is very straight forward and can use the code from the figures \ref{lst.exprImplementation} and \ref{lst.calcJacImplemenation}.
For each operation only the function \emph{derivativeA} is different and therefore a general template file \emph{unaryExpression.tpp} is written.
The template file expects, that the macros \emph{NAME}, \emph{FUNCTION} and \emph{PRIMAL\_FUNCTION} are defined.
They provide the names of the structure, the operator and a function that calls the operator respectively.
From the name of the structure, the file derives the name \emph{gradNAME} for the function that computes the derivative.
In order to make it more convenient to use the template file multiple times, the three macros are undefined at the end of the file.
The implementation of an unary operation is shown in figure \ref{lst.sinImplementation} for the sine.
\begin{figure}
\begin{lstlisting}[style=numbers]
  template<typename R> inline R gradSin(const R& a, const R& result) {
    return cos(a);
  }
  using std::sin;
  #define NAME Sin
  #define FUNCTION sin
  #define PRIMAL_FUNCTION sin
  #include "unaryExpression.tpp"
\end{lstlisting}
\caption{Expression implementation for the sinus function with the \emph{unaryExpression.tpp} template file.}
\label{lst.sinImplementation}
\end{figure}

For binary operators, the same principle can be used, but it requires more predefined functions.
\emph{binaryExpressions.tpp} expects the same predefined macros but also expects the functions \emph{derv(11|11M|10|10M|01|01M)\_NAME} to be defined.
They cover all the cases how a binary operator can be called: with a constant as the first argument, with a constant as the second argument and for all cases when the multiplier is $1.0$ or different.
This requires more effort to implement a new binary operator but gives the developer all possible options for optimizations.
An example implementation for the multiplication is show in figure \ref{lst.mulImplementation}.
\begin{figure}
\begin{lstlisting}[style=numbers]
  template<typename Data, typename R, typename A, typename B>
  inline void derv11_Multiply(Data& data, const A& a, const B& b, const R& result) {
    a.calcGradient(data, b.getValue());
    b.calcGradient(data, a.getValue());
  }
  template<typename Data, typename R, typename A, typename B>
  inline void derv11M_Multiply(Data& data, const A& a, const B& b, const R& result, const R& multiplier) {
    a.calcGradient(data, b.getValue() * multiplier);
    b.calcGradient(data, a.getValue() * multiplier);
  }
  template<typename Data, typename R, typename A>
  inline void derv10_Multiply(Data& data, const A& a, const typename TypeTraits<R>::PassiveReal& b, const R& result) {
    a.calcGradient(data, b);
  }
  template<typename Data, typename R, typename A>
  inline void derv10M_Multiply(Data& data, const A& a, const typename TypeTraits<R>::PassiveReal& b, const R& result, const R& multiplier) {
    a.calcGradient(data, b * multiplier);
  }
  template<typename Data, typename R, typename B>
  inline void derv01_Multiply(Data& data, const typename TypeTraits<R>::PassiveReal& a, const B& b, const R& result) {
    b.calcGradient(data, a);
  }
  template<typename Data, typename R, typename B>
  inline void derv01M_Multiply(Data& data, const typename TypeTraits<R>::PassiveReal& a, const B& b, const R& result, const R& multiplier) {
    b.calcGradient(data, a * multiplier);
  }
  CODI_OPERATOR_HELPER(Multiply, *)
  #define NAME Multiply
  #define FUNCTION operator *
  #define PRIMAL_FUNCTION primal_Multiply
  #include "binaryExpression.tpp"
\end{lstlisting}
\caption{Expression implementation for the multiplication with the \emph{binaryExpression.tpp} template file.}
\label{lst.mulImplementation}
\end{figure}
Macro definitions are avoided in the implementation as much as possible in order to ease debugging.
With the chosen implementation most of the code has a specific line and is not defined in a multi line macro.

\subsection{Design of the tape and calculation type}

The missing ingredients are now the implementation of the calculation types, which the users can use in their software and the tapes that store the Jacobi data.

The calculation type is called \emph{ActiveReal}, which represents the lvalues in a program.
This type has to implement the \emph{Expression} interface and acts therefore as a termination point for the expression templates.
The question is now how much logic the \emph{ActiveReal} structure should contain.
If it contains some logic that is special for a specific tape implementation, then an extra \emph{ActiveReal} implementation is needed for each tape, which would make it quite involved to add new tapes.
We opted for removing all logic from the \emph{ActiveReal} implementation.
It stores only the primal values for the computation, the specific data for the tape and all function calls are forwarded to the tape interface.
Because of this choice no extra interface needs to be defined for the \emph{ActiveReal}.
In addition it defines the regular convenience functions to retrieve the data.

The interfaces for the tapes are separated into one that is called from the \emph{ActiveReal} and an interface for the user interaction.
The interaction with \emph{ActiveReals} requires functions for the creation and destruction of the tape specific data and functions that trigger the storing of the statements.
A statement is stored with the function \emph{store}, here, a tape implementation can access the data from the expression and perform the necessary actions for AD.
The design principle behind CoDiPack is, that the store method calls the \emph{calcGradient} method of the expression.
This causes a recursive call of the \emph{calcGradient} method until the \emph{ActiveReal} types are reached.
As discussed above, we did not want to add any logic to the \emph{ActiveReal} implementation and therefore the call of \emph{calcGradient} is forwarded to the \emph{pushJacobi} method on the tape.
The call hierarchy is then:
\begin{center}
	\begin{tabular}{ccccccc}
	Tape & & Expression & & Active Real& & Tape\\
	store & $\rightarrow$ & calcGradient & $\rightarrow$ & calcGradient & $\rightarrow$ & pushJacobi
	\end{tabular}
\end{center}
which shows that the tape calls itself, through the \emph{calcGradient} method.
In the \emph{pushJacobi} method the tape implementation can therefore gather information about the Jacobian of the expression.

The user interface is more involved and is only required for the reverse mode.
The most important functions are \emph{registerInput}, \emph{registerOutput}, \emph{evaluate}, \emph{setActive} and \emph{setPassive}.
They are needed in order to declare what are the inputs and outputs of the area that is taped.
\emph{evaluate} performs the evaluation of equation \eqref{eq.revOperator} for all the statements in the area that is marked with \emph{setActive} and \emph{setPassive}.

\subsection{Implementation of the tapes}
The implementation of the forward AD mode described in table \ref{tab.ad_fwd} is now very simple.
The full logic is shown in figure \ref{lst.forwardImpl} with a few abbreviations.
All assignment operators will call the \emph{store} method in line 8 from the tape interface.
This method initializes a zero gradient and calls the \emph{calcGradient} method on the expression in order to compute the gradient for the whole statement.
The recursive nature of the \emph{calcGradient} implementation will cause a call of \emph{pushJacobi} for each argument of the statement, which contains the gradient of the statement with respect to this variable.
According to the forward mode in table \ref{tab.ad_fwd}, we only need to multiply this gradient value with the dot value and add it to the dot value of the left hand side, which is implemented in line 18 of figure \ref{lst.forwardImpl}.
Because no external resources are used in the method, the compiler is able to optimize the code as aggressive as possible, which should yield optimal performance results.
\begin{figure}
\begin{lstlisting}[style=numbers]
template<typename R>
  class ForwardEvaluation : public TapeInterface<R, R>{
  public:

    typedef R GradientData;

    template<typename Rhs>
    inline void store(R& value, GradientData& lhsTangent, const Rhs& rhs) {
      R gradient = R();
      rhs.template calcGradient<R>(gradient);
      lhsTangent  = gradient;
      value = rhs.getValue();
    }

    template<typename Data>
    inline void pushJacobi(Data& lhsTangent, const R& jacobi, const R& value, const GradientData& curTangent) {
      ENABLE_CHECK(OptIgnoreInvalidJacobies, isfinite(jacobi)) {
        lhsTangent += jacobi * curTangent;
      }
    }

    ...
  }
\end{lstlisting}
\caption{Tape implementation for the AD forward mode.}
\label{lst.forwardImpl}
\end{figure}

The real challenge is the efficient implementation of the reverse AD mode.
We identified three different data items at the beginning of the section, that we need to store.
There are the number of arguments for each statement, the Jacobi and the index for each argument.
This data is represented in three different data streams, that have two different running indices.
The number of arguments are written per statement and the Jacobi and index are written per argument which makes the these streams run faster than the first one.
In addition we add two data stream for user defined functions and the position where the user defined function needs to be evaluated.
These two streams have a third running index that is even slower, than the one of the arguments per statement.
The management of five different data streams with three different running indices is already quite involved and we anticipate, that other tape developments will require additional data streams with other running indices.
Therefore, we implemented a generalized solution in \emph{chunkVector.hpp}, that can handle an arbitrary set of data streams with different running indices.

The implementation in \emph{chunkVector.hpp} is a recursive data structure which is called \emph{chunk vector}.
It has a template argument for a child vector and uses the child vector's position type to define its own position, therefore every time the position of the parent is queried, the position of the child is also returned.
The same is true when the position of the parent is set (e.g. \emph{reset}), then the same method is called on the child with the child's position.
This has the advantage, that all data streams can be handled as one data object.
All operations are only applied to the root vector and recursively evaluated on all child vectors.
If a new data stream needs to be added, then only a new vector is inserted in the recursive structure and no additional logic needs to be implemented.

The data streams are now stored in chunks of data.
Each chunk can have multiple entries like the \emph{double} for the Jacobi and the corresponding index, which makes it possible to handle data streams with the same running index in one chunk vector.
Because no array of structures is required by this implementation, the memory layout will be optimal for caching and no dead memory is generated because of padding bytes.
This will yield optimal performance on HPC clusters that are optimized for continuous memory access.

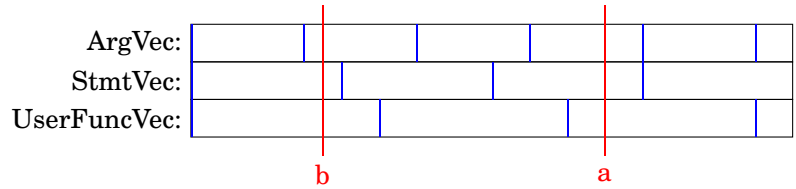
\begin{figure}
	\center

	\begin{tikzpicture}[node distance=1cm]
	  \tikzset{
	    fitting node/.style={
	      inner sep=0pt,
	      fill=none,
	      draw=none,
	      reset transform,
	      fit={(\pgf@pathminx,\pgf@pathminy) (\pgf@pathmaxx,\pgf@pathmaxy)}
	    },
	    reset transform/.code={\pgftransformreset}
	  }
	  \tikzset{
	      mymark/.style={},
	      myarrow/.style={->, >=latex', shorten >=1pt, thick},
	      myarrow2/.style={->, >=latex', shorten >=1pt, thick, dashed},
	      mylabel/.style={text width=7em, text centered}
	  }
	  \newcommand\IndexColor{blue}

	    \draw[black] (0cm,0cm) rectangle (8cm, 0.5cm);
	    \node[left] at (-0cm, 0.25cm) {ArgVec:};

	    \draw[black] (0cm,-0.5cm) rectangle (8cm, 0.0cm);
	    \node[left] at (-0cm, -0.25cm) {StmtVec:};

	    \draw[black] (0cm,-1.0cm) rectangle (8cm, -0.5cm);
	    \node[left] at (-0cm, -0.75cm) {UserFuncVec:};

	    \fill[\IndexColor] (0cm, -0.5cm) rectangle (0.02cm, 0.0cm);
	    \fill[\IndexColor] (2cm, -0.5cm) rectangle (2.02cm, 0.0cm);
	    \fill[\IndexColor] (4cm, -0.5cm) rectangle (4.02cm, 0.0cm);
	    \fill[\IndexColor] (6cm, -0.5cm) rectangle (6.02cm, 0.0cm);

	    \fill[\IndexColor] (0.0cm, -0.0cm) rectangle (0.02cm, 0.5cm);
	    \fill[\IndexColor] (1.5cm, -0.0cm) rectangle (1.52cm, 0.5cm);
	    \fill[\IndexColor] (3.0cm, -0.0cm) rectangle (3.02cm, 0.5cm);
	    \fill[\IndexColor] (4.5cm, -0.0cm) rectangle (4.52cm, 0.5cm);
	    \fill[\IndexColor] (6.0cm, -0.0cm) rectangle (6.02cm, 0.5cm);
	    \fill[\IndexColor] (7.5cm, -0.0cm) rectangle (7.52cm, 0.5cm);

	    \fill[\IndexColor] (0.0cm, -1.0cm) rectangle (0.02cm, -0.5cm);
	    \fill[\IndexColor] (2.5cm, -1.0cm) rectangle (2.52cm, -0.5cm);
	    \fill[\IndexColor] (5.0cm, -1.0cm) rectangle (5.02cm, -0.5cm);
	    \fill[\IndexColor] (7.5cm, -1.0cm) rectangle (7.52cm, -0.5cm);

	    \fill[red] (5.50cm, -1.25cm) rectangle (5.52cm, 0.75cm);
	    \fill[red] (1.75cm, -1.25cm) rectangle (1.77cm, 0.75cm);

	    \node[red] at (1.75cm, -1.5cm) {b};
	    \node[red] at (5.50cm, -1.5cm) {a};
	\end{tikzpicture}
	\caption{Example for the data layout of the Jacobi tape implementation. The blue bars show the chunk boundaries and the marks $a$ and $b$ a possible region for the interpretation.}
	\label{fig:chunkLayout}
\end{figure}

With these implementations the five arrays can be easily defined.
The user functions and the positions for them are always accessed at the same time, therefore they have the same running index.
The chunk is instantiated as \emph{Chunk2<ExtFuncData, ChildPosition>} where \emph{ExtFuncData} is the data for the user defined functions and the \emph{ChildPosition} is the position type of the child vector.
The Jacobi entries and indices for the arguments have also the same running index and can therefore be represented by the chunk \emph{Chunk2<double, int>}, which creates separate vectors for the Jacobies and indices.
The fifth array is the number of arguments for each statement.
It can be represented by the chunk \emph{Chunk1<uint8\_t>}.
With these three chunk definitions the corresponding three chunk vectors \emph{User Function Vector}, \emph{Argument Vector} and \emph{Statement Vector} are created and linked together as
\begin{equation*}
	\text{User Function Vector } \rightarrow \text{Statement Vector} \rightarrow \text{Argument Vector} \rightarrow \text{linear index,}
\end{equation*}
with the linear index as the terminator of the dependency chain.
The linear index just defines a counter for the current statement and has therefore no associated array.
The dependency is now illustrated in figure \ref{fig:chunkLayout}.
It illustrates the data chunks, the vector implementation and the recursive nature.
Each chunk boundary is illustrated by a blue bar.
The lowest lane contains the user function data and has the slowest running index.
Whereas, the mid lane contains the statement data and has a faster running index and the top lane contains the data for each argument and is written therefore most often and has the fastest running index.
The linear index contains no data, it is just used for position calculations.

When the data is evaluated from position a to position b, then 6 chunk boundaries are crossed.
At these points, the interpretation is stopped, the next chunk is loaded and then the interpretation is continued.
Because of the chunk vector design, the chunk management is fully automatic.

There are now several implementations of the reverse AD mode.
The most general one is the \emph{RealReverse} type, which uses a configuration of the chunk vectors that allocates new memory on the fly.
Because of the required bounds checking, this implementation requires at least two if-statements that are evaluated when an operation is stored on the tape.

There is also the \emph{RealReverseUnchecked} type, that is configured such that no bounds checks are performed.
This has the advantage, that no if-statements are required when an operation is stored on the tape and should yield faster code.
The disadvantage of this type is, that the user has to allocate the required memory in advance.

For both types holds, that they are compatible with c-like memory operations (e.g. memcpy).
They can also be configured by various global parameters, that enable or disable additional checks or optimizations.
A short overview of the parameters is given in the results section of this paper.

\subsection{Index reuse}
At the start of the chapter the linear indexing model was introduced.
This model allows the types of the AD tool to be compatible with c-like memory operations, but it produces large adjoint vectors where most of the memory is used just once.
Several things have to be considered for an implementation that reuses indices.

The \emph{ActiveReal} types require now a destructor in order to release the index.
Therefore, an index manager is implemented, that stores the freed indices in a list and tracks the maximum amount of live indices.
A best practice for the index handling has yet to be determined (e.g. use a sorted list, store ranges of free indices), the current implementation simply uses a stack for the index management in order to be as fast as possible.
If e.g. a sorting logic is used in the index handling, then it impacts the evaluation of each statement which has a huge performance impact on the runtime.

Furthermore, because the indices are reused, every \emph{ActiveReal} needs a distinct index.
If two \emph{ActiveReal}'s  would have the same index, it would be released twice and can then be given to two different variables which would use the same memory location in the reverse evaluation.
With this in mind, the \emph{ActiveReals} with an index reuse technique can no longer support c-like memory operations.

The advantage of the index reuse is the reduced size of the adjoint vector.
Instead of having the size of the total amount of variables, it shrinks to the size of the maximum amount of used variables.
Because of the reduced size, the reverse evaluation is usually faster.

The CoDiPack types with an index reuse have a \emph{Index} suffix in there type names.
This results into the two new types \emph{RealReverseIndexUnchecked} and  \emph{RealReverseIndex}.

\section{Tests}\label{sec.test}

\subsection{Coupled burgers equation}

The coupled burgers equation \cite{biazar2009exact,bahadir2003fully,zhu2010numerical} is chosen as lightweight test, that can be used to do a rapid evaluation how some changes in CoDiPack affect the performance.

The equations

\begin{align}
	u_t + uu_x + vu_y &= \frac{1}{R}(u_{xx} + u_{yy}) \\
	v_t + uv_x + vv_y &= \frac{1}{R}(v_{xx} + v_{yy})
\end{align}

are discretised with an upwind finite difference scheme. The initial conditions are:
\begin{align}
	u(x, y, 0) &= x + y \quad (x,y) \in D \\
	v(x, y, 0) &= x - y \quad (x,y) \in D
\end{align}

and the  exact solution is (\cite{biazar2009exact})
\begin{align}
	u(x, y, t) &= \frac{x + y - 2xt}{1 - 2t^2} \quad (x,y,t) \in D \times \R\\
	v(x, y, t) &= \frac{x + y - 2xt}{1 - 2t^2} \quad (x,y,t) \in D \times \R \eqdot
\end{align}

The computational domain $D$ is the unit square $D = [0,1] \times [0,1] \subset \R \times \R$ and the boundary conditions are taken from the exact solution.
As far as the differentiation is concerned, we chose as the input parameters the initial solution of the time stepping scheme and as the output parameter we take the norm of the final solution.
%The computed sensitivities describe how much the final solution depends on the initial solution, which resembles a parameter fitting problem.
%It is not that important which derivatives are computed, because this test case is only used for a rapid evaluation of changes in CoDiPack.

For the implementation of the program, all methods are programmed in such a way that they can be inlined and the requested memory is allocated and initialized before the time measurement starts.
Especially the required memory for the tape is computed beforehand and allocated.
This yields a very stable time measurements which are run on one node of the Elwetritsch cluster at the TU Kaiserslautern, which consists of two Intel E5-2670 CPU's with a total of 16 cores and 128 GB of Memory.
The general setup calculates the burgers equations on a 601x601 grid with 32 iterations and the time measurement is repeated 10 times.
For the time measurements two different configurations are tested:
\begin{itemize}
	\item The \emph{multi} test configuration runs the same process on each of the 16 cores.
		This setup simulates a use case where the full node is used for computation and every core uses the memory bandwidth of the socket.
	\item The \emph{single} test configuration runs just one process on the whole node.
		This eliminates the memory bandwidth limitations and provides a better view on the computational performance.
\end{itemize}

Both test configurations will be evaluated with the default CoDiPack settings and then with special configurations options in order to see how these options affect the performance of CoDiPack.

\subsection{SU2}
An important application where derivatives are nowadays frequently needed is numerical optimization. When constraints are defined using partial differential equations (PDE), this usually requires efficient adjoint methods. In that case AD facilitates the development of those solvers in the discrete setting. This is especially valuable in computational fluid dynamics, where the analysis, i.e. the evaluation of the constraining PDE, is achieved by highly complex algorithms.

Recently, \emph{CoDiPack} was applied by the authors to the open-source framework \emph{SU2} \cite{economon2015su2}. The latter is a collection of tools for the analysis and optimization of internal and external aerodynamic problems using a Finite-Volume method. The differentiated code was used to generate a flexible and robust discrete adjoint solver for the Reynolds Averaged Navier-Stokes equations \cite{Albring2015} optionally coupled with the Ffowcs-Williams-Hawkings equation \cite{Zhou_etal_2016b} for aeroacoustic optimizations. The adjoint solver is based on the fixed-point formulation of the underlying solver for the discretized state equation \cite{Christianson1994}. That is, we assume that feasible solutions $U^*$ of the state equation
\begin{eqnarray}\label{eq.state_equation}
U = G(U,X)
\end{eqnarray}
are computed by the iteration $U^{n+1} = G(U^n,X)$ for $n \rightarrow N^*$. $X$ represents the design variables and $G(U)$ some (pseudo) time-stepping scheme like the explicit or implicit Euler method. By applying the first order necessary conditions on the optimization problem to minimize some scalar objective function $J(U,X)$ with the constraint that the state equation \eqref{eq.state_equation} is fulfilled, we end up with the following fixed-point equation for the adjoint state $\bar{U}$:
\begin{eqnarray}\label{eq.adjstate_equation}
\bar{U} = \left[\frac{\partial J(U^*,X)}{\partial U}\right]^T +  \left[\frac{\partial G(U^*,X)}{\partial U}\right]^T \bar{U}
\end{eqnarray}
This equation can be solved using the iterative scheme
\begin{eqnarray}\label{eq.adjstate_scheme}
\bar{U}^{n+1} = \left[\frac{\partial J(U^*,X)}{\partial U}\right]^T +  \left[\frac{\partial G(U^*,X)}{\partial U}\right]^T \bar{U}^n, \quad \text{for}~~~ n\rightarrow \bar{N}^*.
\end{eqnarray} 
The above scheme can be easily constructed by applying AD to the code that computes $J$ and $G$. It is important to note, that since we need the right hand side of equation \eqref{eq.adjstate_scheme} repeatedly at the fixed-point $U^*$, we only need to tape the Jacobians of the statements once at the last fixed-point iteration. Hence the time required for the interpretation determines the overall run-time. In contrast to other implementations of adjoint solvers, where AD is only applied to certain part of the code, we replaced all occurences of the default computation type with the AD-type provided by CoDi. Although this leads to a small overhead in run time (one \textit{if}-statement per expression), the maintainability and on-the-fly differentation of new features typically outweigh this disadvantage. 

As the testcase we consider the inviscid flow over the LM1021 supersonic aircraft. The computational mesh consists of $5{,}730{,}841$ interior elements and the aircraft is discretized using $214{,}586$ boundary elements. For the spatial integration we use a central scheme with artifical dissipation in combination with an explicit Euler method for the pseudo-time stepping.

\section{Results}

\subsection{Coupled burgers equation}\label{sec.burgers}
\begin{figure}
  \center
  \includegraphics{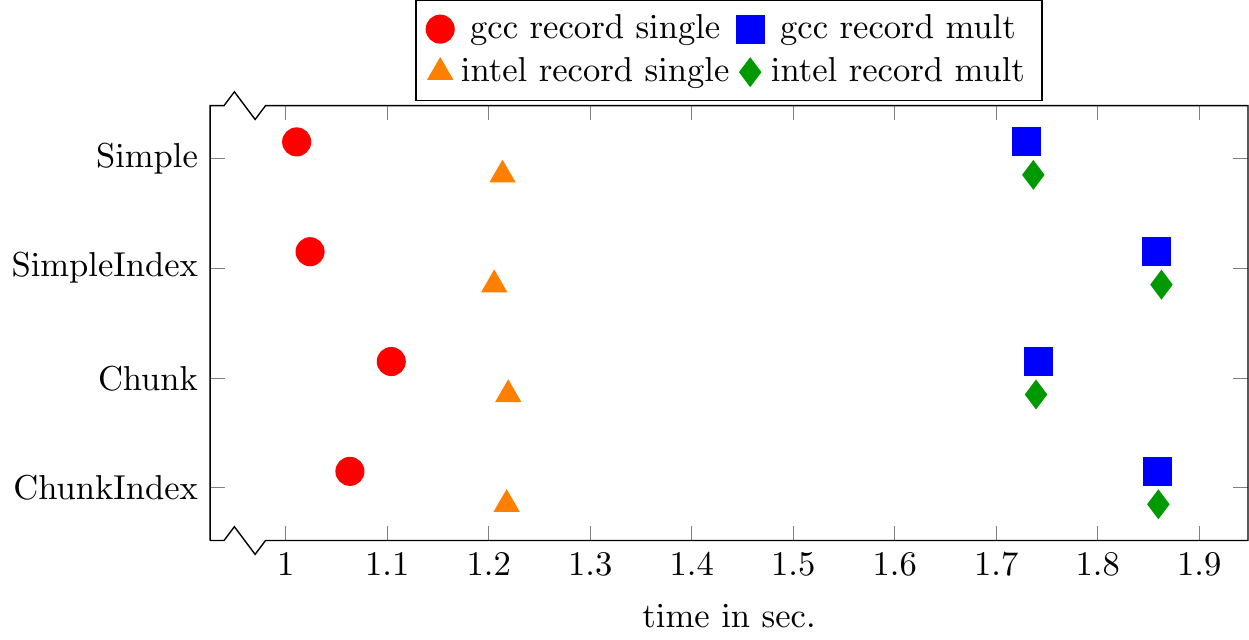}

  \caption{CoDiPack recording times for the coupled burgers equation test case.
    The circles display the averaged time over all runs and the bars show the minimum and maximum time values.}
  \label{fig:commonRecord}
\end{figure}

The first analysis discusses the general timings of the implementation and the comparison between the Intel and gcc compiler.
The recording times for the coupled burgers equation are always averaged over all processors and runs and are shown in figure \ref{fig:commonRecord}.
Every circle in the plot shows the averaged time and the bars show the minimum and maximum times.
The results are shown for the Intel and gcc compiler in the single and multi configuration.
For each compiler and configuration the four different available tapes are checked.
These correspond to the \emph{RealReverseUnchecked} (Unchecked), the \emph{RealReverseIndexUnchecked} (UncheckIndex), the \emph{RealReverse} (Chunk) and the \emph{RealReverseIndex} (ChunkIndex) types of CoDiPack.

As we can see, there is quite some discrepancy between the singe and the multi variant of the burgers test case.
The single variant runs approximately 80\% faster than the multi variant, because of the memory bandwidth limitation of the computing node.
In the single test case only one process uses the full bandwidth, while in the multi test case 16 processes share the memory bandwidth.
This has some interesting additional effects on the run time for the Intel and gcc compiler, as they differ for the single configuration, which is no longer seen in the multi case.
This shows that a certain extend of compiler and implementation differences, can be hidden in the memory bandwidth.

It can also be seen that for the single test case the required recording time increases with the complexity of the tape implementation.
The Unchecked tape is the simplest one and only a small increase is seen for the UncheckIndex version, that additionally uses an index handler.
The jump from the Unchecked variants to the Chunk variants, which allocate memory on the fly and perform bound checking, is around 10\% computation time.
Interestingly, the index reuse is faster than the linear indexing in that case, probably due to caching effects.
In the multi test case all these effects are no longer visible.
The UncheckedIndex and ChunkIndex types require more taping time because they need to store additional data.

\begin{figure}
  \center
  \includegraphics{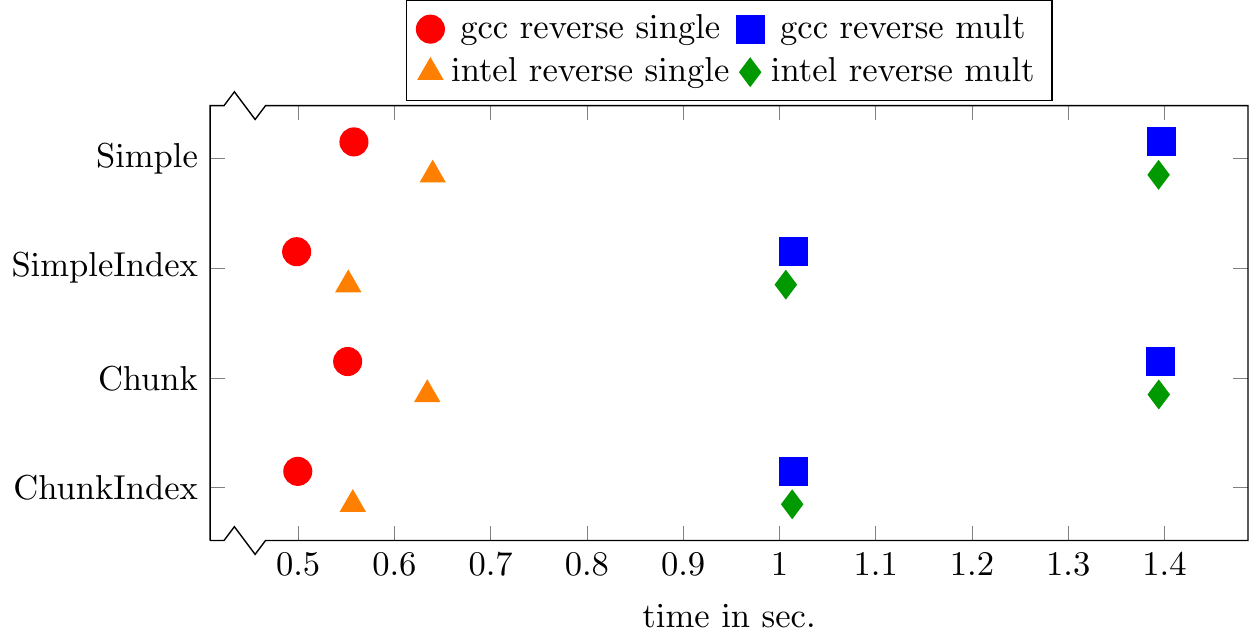}

  \caption{CoDiPack reverse interpretation times for the coupled burgers equation test case.
    The circles display the averaged time over all runs and the bars show the minimum and maximum time values.}
  \label{fig:commonReverse}
\end{figure}

The interpretation time for the burgers equation is shown in figure \ref{fig:commonReverse}.
Here, the difference between the Unchecked and Chunk types is quite small, because the code for both tape types is nearly the same.
Only the index reuse types show a different timing.
They require less memory in the reverse interpretation and therefore they are faster in the interpretation.
The Intel compiler is again a little bit slower in the single test case, which is no longer seen in the multi case with the memory bandwidth limitation.
Figure \ref{fig:commonReverse} also shows, that the reduced memory of the Index tapes have a large impact on the reverse interpretation time.
We can therefore reach the conclusion, that everything which saves memory can also improve the required time for the AD process.

As was already shown in \cite{hogan2014fast}, the performance of AD not only depends on the structure of the code and its memory footprint, but also on the number of transcendental functions. Usually the compiler can aggressively optimize code that contains only non-transcendental functions, but the application of AD interferes in general with this optimizations. Hence the more transcendental functions occur in the code, the better the relative performance of recording and interpretation. Since the algorithm for the Burgers equation represents an extreme case with no transcendental functions at all, we can except relatively high numbers when comparing the runtime of the derivative computation with the runtime of the primal code (indeed we get a factor $22$ and $26$ for the single and mult case, respectively).

The next analysis considers the default block size for the chunk tapes.
The data in table \ref{tab.blockSizes} shows that the recording of the tape is not influenced by the size of the blocks.
This is however different when the evaluation of the tape is considered.
Here, the multi test configuration shows only an overhead for very small blocks.
On the contrary, for the single case a large influence is seen on the block size up to 0.5 million entries, which is equivalent to a size of 6 MB for 12 byte entries.
A small increase in time is also seen for blocks with more than 33 million entries.
The reason why the influence is only seen in the tape interpretation of the single test case is that for the other cases the memory bandwidth is the dominating factor.
In order to have an optimal block size for all cases the default is set in CoDiPack to 2 million entries.

\begin{table}
	\tbl{Block size time comparison in seconds for the coupled burgers equation\label{tab.blockSizes}}{%
  \begin{tabular}{|l|c|c|c|c|}
	  \hline
	  Block Size & Record multi & Record single & Interpret multi & Interpret single\\
    \hline
            1,024 & 2.15 & 1.30 & 1.89 & 0.98 \\
            2,048 & 2.15 & 1.29 & 1.81 & 0.85 \\
            4,096 & 2.16 & 1.28 & 1.77 & 0.79 \\
            8,192 & 2.17 & 1.27 & 1.75 & 0.74 \\
           16,384 & 2.16 & 1.27 & 1.75 & 0.73 \\
           32,768 & 2.18 & 1.28 & 1.74 & 0.71 \\
          131,072 & 2.18 & 1.28 & 1.74 & 0.71 \\
          262,144 & 2.18 & 1.29 & 1.74 & 0.72 \\
          524,288 & 2.16 & 1.28 & 1.75 & 0.70 \\
        1,048,576 & 2.16 & 1.29 & 1.75 & 0.72 \\
        2,097,152 & 2.16 & 1.28 & 1.75 & 0.70 \\
        4,194,304 & 2.16 & 1.28 & 1.75 & 0.70 \\
        8,388,608 & 2.16 & 1.28 & 1.75 & 0.70 \\
       16,777,216 & 2.16 & 1.28 & 1.75 & 0.70 \\
       33,554,432 & 2.16 & 1.28 & 1.75 & 0.71 \\
       67,108,864 & 2.16 & 1.28 & 1.75 & 0.72 \\
      134,217,728 & 2.17 & 1.28 & 1.75 & 0.73 \\
    \hline
  \end{tabular}
  }
\end{table}

The implementation of CoDiPack contains several switches that can be used to fine tune the tapes to the needs of the user.
The data in table \ref{tab.switches} shows the switches separated into the ones that influence the recording of the tape and the ones that influence the interpretation of the tape:
\begin{itemize}
	\item With the \emph{Check expression arguments} switch enabled, CoDiPack performs checks if the arguments for the elemental functions are in the differentiable domain.
	\item \emph{Ignore invalid Jacobians} and \emph{Ignore zero Jacobians} prevent zero and invalid Jacobians to be recorded on the tape.
	\item \emph{Check tape activity} enables the test if the tape is currently active and should therefore record statements.
	\item The reverse switch \emph{Skip zero adjoints} prevents CoDiPack from performing the update $ \bar a \aeq c * 0.0$ in the reverse interpretation.
\end{itemize}
Because the test case is highly optimized, only the ignoring of zero Jacobies changes the tape, but to margin, that can be ignored.
The memory bandwidth limited case shows only a small increase in the recording and evaluation time when all switches are enabled because most of the logic can be hidden in the memory latency.
Therefore, for large cases all checks can be enabled and the additional time will be minimal.
The single case shows an increase of 10\% in the computation time for the recording and evaluation of the tape.
This shows, that if CoDiPack is used on very small portion of a code, where the tape size is very small, it can improve the performance when specific checks are disabled.

\begin{table}
	\tbl{Configuration switches time comparison in seconds for the coupled burgers equation\label{tab.switches}}{%
  \begin{tabular}{|l|c|c|c|c|}
	  \hline
	  Record switch & Record multi & Record single & Interpret multi & Interpret single\\
    \hline
                       All off & 2.13 & 1.12 & 1.75 & 0.67 \\
    Check expression arguments & 2.13 & 1.12 & 1.76 & 0.67 \\
      Ignore invalid Jacobian's & 2.15 & 1.15 & 1.74 & 0.67 \\
         Ignore zero Jacobian's & 2.14 & 1.15 & 1.76 & 0.71 \\
           Check tape activity & 2.14 & 1.16 & 1.76 & 0.71 \\
                        All on & 2.15 & 1.23 & 1.76 & 0.75 \\
    \hline
	  Interpretation switch & Record multi & Record single & Interpret multi & Interpret single\\
    \hline
                       All off & 2.13 & 1.12 & 1.75 & 0.67 \\
            Skip zero adjoints & 2.13 & 1.13 & 1.76 & 0.71 \\
                        All on & 2.15 & 1.23 & 1.76 & 0.75 \\
    \hline
  \end{tabular}
  }
\end{table}
	
\subsection{SU2}

\begin{figure}[!ht]
	\subfloat[Relative time values.\label{subfig:fig:lm1021_time}]{%
  		\includegraphics[width=0.45\textwidth]{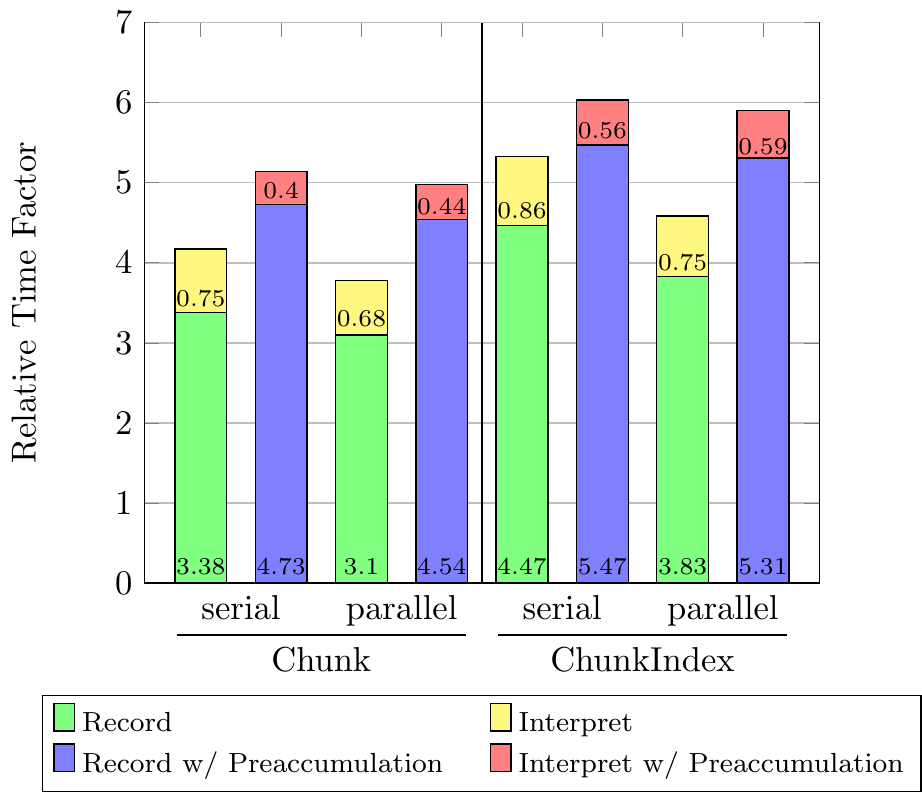}
  	}
	\subfloat[Relative memory values\label{subfig:fig:lm1021_memory}]{%
  		\includegraphics[width=0.405\textwidth]{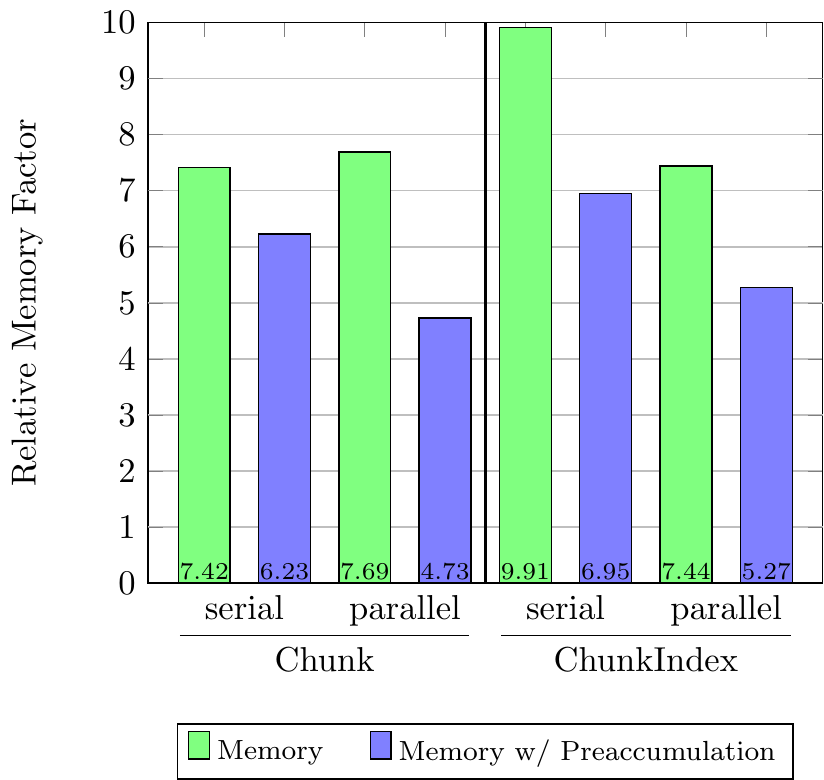}
  	}
  \caption{Measurements of the SU2 CFD suite for the LM1021 aircraft. All values are give as relative measurements with respect to the unmodified primal code.}
	\label{fig:lm1021}
  \end{figure}

Figure \ref{fig:lm1021} shows the relative time and memory factors for 100 iterations of equation \eqref{eq.adjstate_scheme} with respect to one evaluation of equation \eqref{eq.state_equation}.
Here we also distinguish between the serial run and the parallel run utilizing a full compute node with 16 cores and the Chunk and ChunkIndex tapes. Due to its structure, CoDi allows the direct access to the Jacobian values stored on the tape. This enables the use of advanced AD techniques like the preaccumulation of Jacobians \cite{Utke2005FBB} where parts of the tape are already interpreted during the recording. In SU2 this method is employed in certain parts of straight-line code to significantly reduce the memory and interpretation time at the cost of an increased recording time. We include this approach in the discussion, since it represents the common setting when using the adjoint solver.

First of all there is no performance degradation noticeable for the parallel computations compared to the serial computations. This indicates that the latter is already bandwidth limited. In fact, the parallel case even consistently shows a slightly lower factor.
Although the majority of operations in SU2 are multiplications and additions, in some parts of the code also transcendental functions are used which intervene with compiler optimizations. Indeed, unlike the test case presented in the previous subsection, we get relative run-time factors between 3.1 to 4.47 for the recording and between 0.4 to 0.86 for the interpretation of one flow iteration without preaccumulation. Again we emphasize that this represents a "black-box" differentiation that contains \emph{all} routines to compute the flow update in equation \eqref{eq.state_equation}, even if they do not contribute to the final gradient and can therefore be considered as passive. Hence little knowledge of the code is required in that case. The application of preaccumulation requires slightly more knowledge of the code to identify inputs/outputs of certain regions. Still, once identified, the modifications to apply the method are minor. Figure \ref{fig:lm1021} also shows the effect of preaccumulation on the run-time and memory. Although the time for taping increases by roughly $40\%$, the interpretion time reduces by almost $50\%$, which in turn reduces the overall run-time of the adjoint solver by nearly $50\%$. In addition the memory consumption is also reduced by a great amount.

In contrast to the Burgers test case in section \ref{sec.burgers}, here the non-linear indexing scheme (represented by the type \emph{ChunkIndex}) does not offer a better performance compared to the \emph{Chunk} type with linear indexing. A detailed analysis showed that this is due to a high amount of "active copies" in SU2, i.e. copy/assignment operations with just one active type on the right hand side. For the linear indexing a simple copy including the index value is sufficient in that case. For the non-linear indexing type a more complex treatment is necessary since each index needs to be unique in the application.
Therefore a statement has to be written for each copy operation.

%The comparison between the Chunk and ChunkIndex tapes provides similar insights as for the burgers equation.
%Due to the memory overhead in the tape recording, the recording time is slower.
%Usually the reverse interpretation time should be faster, because of the reduced adjoint vector size and therefore a reduced %memory size.
%But the memory plot \ref{subfig:fig:lm1021_memory} shows that not much memory is saved.
%The index reuse tape can not perform several optimizations for assign statements, that are otherwise possible.
%Therefore, the assign statements need to be recorded which increases the tape size.
%There are some options to circumvent this deficit of the index reuse technique, but there are not yet implemented in SU2.

\section{Conclusion and Outlook}
In this paper we demonstrated the efficient implementation of a novel AD tool that employs Jacobi taping with expression templates. Unlike other tools the layout of CodiPack is designed in such a way that other taping approaches can be easily added to further enhance its capabilities, while still maintaining an easy to use interface and high performance.
Furthermore, the recursive layout of the data streams allow for a simple addition of new streams and several design choices such as the avoidance of preprocessor macros make the code easy to understand.
CoDiPack provides a useful basis that hopefully encourages people to join the currently stagnating research in implementation strategies for AD.

The performance of the Expression Template approach heavily relies on the capability of the compiler to inline certain routines. Hence to assure that CoDiPack offers a compiler independent behavior we first compared the performance between the Intel and gcc compiler. Here it was shown that both compilers give indeed similar performance, though the gcc compiler was slightly faster when just a single process is used. This difference diminishes however when occupying all available cores of a node because the memory bandwidth becomes the limiting factor. Similar results were reported for the indexing and memory management schemes for both compilers. Using the same compiler, it was shown that there is only a moderate performance degradation during the recording when the more complex chunk memory management implementation is used compared to the simple management without any bound checking. The former offers a more user friendly behavior without the need to know the tape size a-priori. Furthermore the types that use an index handler need less run-time for the interpretation due to the reduced tape size.

The application to the CFD software SU2 certified CoDiPack's high performance even for the case of a "black-box" differentiation of complex code and in parallel. High-level tuning using preaccumulation techniques can help to further decrease the memory consumption.

Future research is devoted to the improvement of the performance using novel indexing and memory management schemes. We will also expand the research to the field of analysis of codes, to provide guidelines which CoDi type is optimal for what kind of code structure. Additionally the user interface in CoDiPack is constantly extended to provide convenient routines to easily handle user differentiated functions or preaccumulation.

\bibliographystyle{ACM-Reference-Format-Journals}
\bibliography{citations}

% History dates
\received{MM YYYY}{MM YYYY}{MM YYYY}
\end{document}